\documentclass[12pt]{article}
\usepackage{graphicx,amsmath,amssymb}
\newcommand{\be}{\begin{equation}}
\newcommand{\ee}{\end{equation}}
\newcommand{\ba}{\begin{eqnarray}}
\newcommand{\ea}{\end{eqnarray}}

\def\gl#1{(\ref{#1})}
\def\tr{{\rm tr}}

\def\reg{{\tt reg}}
\def\Dirac{i\partial\!\!\!\!/}

\begin{document}
\centerline{\large\bf On the consistency of Lorentz invariance violation in QED}
\centerline{\large\bf  induced by fermions
in constant axial--vector background}
\bigskip
\bigskip
\renewcommand{\thefootnote}{\fnsymbol{footnote}}
\centerline{\large J. Alfaro$^{1}$, A.A. Andrianov$^{2,}$\footnote{Corresponding author. 
E-mail:  andrianov@bo.infn.it}, M. Cambiaso$^{1}$,}
 \centerline{\large
P. Giacconi$^{3}$ and  R. Soldati$^{3}$}
\bigskip

 \noindent
$^{1}$Facultad de F\'{\i}sica, Pontificia
Universidad Cat\'olica de Chile,
Macul, Santiago de Chile,
Casilla 306, Chile\\
$^{2}$V.A. Fock Department of Theoretical Physics, St. Petersburg State University,
 198504 Sankt-Petersburg, Russia\\$^{3}$Dipartimento di Fisica,
Universit\'a
di Bologna \& INFN, Sezione di
Bologna, 40126 Bologna, Italia\\

\date{}

\begin{abstract}
We show for the first time that the induced parity--even Lorentz invariance
violation can be unambiguously calculated in the physically justified and
minimally broken dimensional regularization scheme, suitably tailored for a
spontaneous Lorentz symmetry breaking in a field theory model.
The quantization of the Lorentz invariance violating quantum
electrodynamics is critically  examined and shown to be consistent
either for a light--like cosmic anisotropy axial--vector or for
a time--like one, when in the presence of a bare photon mass.
\end{abstract}



\section{Introduction}
The Lorentz and CPT invariance violation (LIV) in quantum electrodynamics
\cite{1,2} has not yet been seen \cite{shore} but, in principle, it might arise in
several ways reviewed in \cite{rev1,rev2}. In particular, spontaneous symmetry
breaking \cite{lsb} may cause LIV after condensation of massless axion--like fields
\cite{ansol,arka} and/or of certain vector fields \cite{veccon} (maybe, of gravitational origin \cite{mukho}),
 as well as
short--distance space--time asymmetries may come from string \cite{string} and
quantum gravity effects \cite{quangrav,alfaro} and non--commutative structure of
the space--time \cite{noncom}.
Whereas the empirical parameterization of LIV  does not represent a tedious and
difficult task \cite{2,shore}, the consistent unraveling of its dynamical origin
is far more subtle and involved.

If LIV occurs spontaneously in QED, due to some vector--like condensate,
then the related low--energy effective action is actually
dominated by the classical
gauge invariant Maxwell--Chern--Simons modified
electrodynamics \cite{1}, with a Chern--Simons (CS)
fixed vector $\eta_{\,\mu}\,.$ In addition,
the low--energy effective action
has to be indeed supplemented by a
Lorentz--invariant bare photon mass $\mu_{\gamma}$ and take
into account  the contribution from the one--loop radiative corrections
\cite{CSrad,AGS,bonn} induced by the fermion sector,
in which some constant axial--vector
$b_{\mu} = \langle B_\mu\rangle= \langle \partial_\mu \theta\,\rangle$
does appear.
The latter one might represent the vacuum expectation value of a vector field
$B_\mu (x)\,,$ such as some torsion field of a cosmological nature,
or of a gradient of some axion field
or quintessence field $\theta(x)\,,$
or anything else.
Whatever it is, it turns out to be  responsible of the Lorentz invariance violation
and, correspondingly, this model will be henceforth referred to as
Lorentz invariant violating quantum electrodynamics (LIVQED).

If LIV manifests itself as a fundamental phenomenon in the large--scale Universe,
it is quite plausible that  LIV is induced universally by different species of
fermion fields coupled to the very same axial--vector $b^{\,\mu}\,,$
albeit with different magnitudes depending upon flavors.
Then both LIV vectors become
\cite{CSrad,AGS} collinear, {\it i.e.} $\eta^{\,\mu} = \zeta b^{\,\mu}\,.$
Meantime it has
been found \cite{ansol,ASS,adam,koslehn} that a consistent quantization
of photons just requires the CS vector to be space--like, whereas for the
consistency of the spinor free field theory a
space--like axial--vector $b^{\,\mu}$ is generally not allowed but for
the pure space--like case which, however, is essentially ruled out
by the experimental data \cite{DP}.

In our Letter the quantization of the Lorentz invariance violating quantum
electrodynamics is critically  examined and shown to be consistent
either for a light--like cosmic anisotropy axial--vector or for
a time--like one, when in the presence of a bare photon mass. 
To this purpose, we show for the first time that the induced parity--even Lorentz invariance
violation can be unambiguously calculated in the physically justified and
minimally broken dimensional regularization scheme, suitably tailored for a
spontaneous Lorentz symmetry breaking in a field theory model.

Two main paths will be focused, along which the incompatibility
in the LIV quantization of the photon and fermion fields
can be actually removed. A first road towards the quantum intrinsic consistency
of LIVQED is opened by the light--like vector $b^{\,\mu}\,.$
As well, the cure can be owed to a radiatively generated mass
for the photon.
The relevant LIVQED Lagrange density in the photon sector reads
\begin{eqnarray}
&&{\cal L}_{\,\gamma}= -\,{\frac14} \left(1 + \xi\,\frac{b^{\,2}}{m_e^2}\right)
F^{\,\nu\lambda}F_{\nu\lambda} +
\xi\,\frac{b_\nu b^{\,\rho}}{2m_e^2}\,F^{\,\nu\lambda} F_{\rho\lambda}\nonumber\\
&& -\,
\frac12\,\zeta\,b_\nu A_\lambda\widetilde
F^{\,\nu\lambda} +\, \frac12\,m^2_\gamma\,A_\nu A^\nu + B\,\partial_\nu A^\nu\, ,
\label{2.1}
\end{eqnarray}
with the photon mass assembling both a bare and an induced one,
$m^2_\gamma=\mu^2_\gamma +\delta m^2_\gamma$ .
As the gauge invariance is broken by the photon
mass, we have suitably introduced the St\"uckelberg's
auxiliary field $B(x)\,,$ together with the usual dual field
tensor $\widetilde F^{\,\nu\lambda}\equiv
(1/2)\,\epsilon^{\nu\lambda\rho\sigma}F_{\rho\sigma}$
and the electron mass $m_e\,.$

As it will be shown in next section, the induced photon mass squared
$\delta m^2_\gamma$
is of order $\alpha\,b_\nu b^\nu \equiv \alpha b^2\,,$  $\alpha = e^2/4\pi$ being
the fine structure constant. The electric charge $e$ is included also into
the induced dimensionless constants $\xi \sim \alpha$ and $\zeta\sim \alpha$
and the related LIV vertices are estimated to be indeed very small
\cite{1,shore} --
see the discussion and concluding remarks in the last section.

\section{ Induced coupling constants}
 Let us now evaluate the constants
$\xi$ and $\zeta$
as well as the  photon mass $m_\gamma\,$ when they are induced by fermions
in an axial background of a cosmological origin.
To this aim, one has
to compute the one--loop induced effective
action from the classical spinor Lagrange density for a
given fermion species $f$ of electric charge $eq_f$: namely,
\begin{eqnarray}
{\cal L}_f={\overline \psi}_f
\Big(\Dirac - m_f - b_f^\mu\,\gamma_\mu\gamma_5
+ eq_f\, A\!\!\!\!/\,\Big)\psi_f\,,
\label{4.5}
\end{eqnarray}
which leads to the Feynman's propagators in
the four dimensional momentum space,
\ba
&& S_f(p)= i\,\Big( \gamma^\nu p_\nu + m_f
 + b^{\,\nu}_f\,\gamma_\nu\gamma_5\Big)\ \times
\label{propagator}\\
&&\times\ \frac{p^2+ b_f^2-m_f^2+2\left(p\cdot b_f +
m_f\,b_f^\lambda\,\gamma_\lambda\right)\gamma_5}{
\left(p^2+b_f^{\,2} - m_f^2+i\varepsilon\right)^2-4
\left[\,(p\cdot b_f)^2-m_f^2\,b_f^2\,\right]}\,.\nonumber
\ea
The one--loop vacuum polarization tensor is formally determined to be
\begin{eqnarray}
\Pi^{\,\nu\sigma}_2(k\,;b_f,m_f) = - ie^2 q_f^2
\int\frac{d^{4}p}{(2\pi)^{4}}\
\tr\left\{\gamma^\nu\,
S_f(p)\,\gamma^\sigma\, S_f(p-k)\right\}\,.
\label{polarM}
\end{eqnarray}

The above formal expression for $\Pi^{\,\nu\sigma}_2$ exhibits, by  power
counting, ultraviolet divergencies. In the presence of LIV due to the background
axial--vectors $b_f^\mu\,,$ the physical cutoff in fermion momenta does emerge
as a result of
fermion--antifermion pair creation at very high energies.
To this concern, it has been proved  \cite{AGS} that the calculations with such
a physical cutoff do actually give the same results of a
Lorentz invariance violating dimensional regularization
scheme (LIVDRS)
$$
\int\frac{d^{4}p}{(2\pi)^{4}}\ \longrightarrow\  \mu^{4-2\omega}\
\int\frac{d^{2\omega}p}{(2\pi)^{2\omega}}
$$
suitably tailored in order to strictly preserve the residual Lorentz symmetry.
This LIVDRS coincides with the conventional one, with
't Hooft--Veltman--Breitenlohner--Maison algebraic rules for gamma--matrices,
when it is applied to the integrand (\ref{polarM}) with fermion
propagators (\ref{propagator}) having the spinor matrices in the numerator.
The general structure of the regularized polarization tensor  turns out to be
\begin{equation}
\reg \Pi^{\,\nu\sigma}_2\ =\ \reg \Pi^{\,\nu\sigma}_{2,\,{\rm even}}\ +\
\reg \Pi^{\,\nu\sigma}_{2,\,{\rm odd}}\,.
\label{4.8}
\end{equation}
The regular odd part has been unambiguously  evaluated in
\cite{AGS} with the help of LIVDRS and for small $| b^\mu | \ll m_e$ reads
\begin{equation}
\reg \Pi^{\,\nu\sigma}_{2,\,{\rm odd}} =
\left(\frac{\alpha}{\pi}\right)\,
2i\,\epsilon^{\,\nu\sigma}_{\quad\rho\lambda}\,k^{\,\lambda}
\sum_f q^2_f\,b_f^{\,\rho}\,,
\label{regodd}
\end{equation}
so that we can eventually identify
\be
\zeta\,b^{\,\mu} = 2\left(\frac{\alpha}{\pi}\right)
\sum_f q^2_f\,b_f^{\,\mu}\,. \label{zeta}
\ee
In the case of a universal Lorentz symmetry
breaking, that means the very same axial--vector $b^{\,\mu}$ for all fermion species,
one can derive
\be
\zeta = 2q^2\left(\frac{\alpha}{\pi}\right)\,,
\ee
where $\sum_f q^2_f\equiv q^2$ is a sum over the normalized fermion electric charges,
$q^2 = 3\cdot3\cdot (1/9 + 4/9) + 3 = 8$ being the result
for three generations of quarks and leptons in the standard model.

With the help of the LIVDRS \cite{AGS}, the even part
of the vacuum polarization tensor can be also found unambiguously.
In this work we focus our attention on
the LIV deviations of free photons on mass shell $k^2 \sim 0$.
The latter ones are expected to be really small
$\Delta k^2 \ll m^2_e$ and therefore it makes sense to retain only leading orders in $k^2$ and  $b^\mu$ .
Correspondingly this part of the polarization tensor  takes the form,
\begin{eqnarray}
\reg \Pi^{\,\nu\sigma}_{2,{\rm even}}\ =
(k^2 g^{\nu\sigma} - k^\nu k^\sigma )\,\Pi_{\,\rm div}
+\ \frac{2\alpha}{3\pi} \sum_f q^{\,2}_f \Biggl(b_f^2\, g^{\nu\sigma}
 - m_f^{-2}\,S_f^{\,\nu\sigma}\Biggr)\,,
\end{eqnarray}
in which we have set
\begin{eqnarray}
S_f^{\,\nu\sigma}
\equiv g^{\nu\sigma}\Bigl[\,(b_f\cdot k)^2 - b_f^2\,k^2\,\Bigr]
- (b_f\cdot k) \left(b_f^\nu\,k^\sigma + b_f^\sigma\,k^\nu\right)
+  k^2\,b_f^\nu\,b_f^\sigma +  b_f^2\, k^\nu k^\sigma\,.
\end{eqnarray}
Here the first term $\Pi_{\,\rm div}$ is logarithmically divergent  and does
renormalize the electric charges in a conventional way, whereas the second term
is finite and involves the sum  over the charged fermions of the standard model.
Thus, for a universal $b^\mu$ the induced constants
in the lagrangian density are eventually determined as follows:
\ba
&&\delta m^2_\gamma =\  -\ \frac{2\alpha}{3\pi}\,\sum_f q_f^2\,b_f^2 \longrightarrow
-\ \frac{16\alpha}{3\pi}\,b^2\, \quad \mbox{\rm for SM}\, ,\\
&& \xi = \frac{2\alpha}{3\pi}
\sum_f q^2_f\left(\frac{m_e}{m_f}\right)^2\ \simeq\
\frac{2\alpha}{3\pi}\, ,
\label{param}
\ea
as the electron is the lightest charged particle.

\section { LIV dispersion law}
Let us now analyze the modified Maxwell's equations
\begin{eqnarray}
&&\left(1+ \xi\,\frac{b^{\,2}}{m_e^2}\right)\partial_\lambda F^{\,\lambda\nu} -
\frac{\xi}{m_e^2}\left(b^{\,\rho} b_\lambda \partial_\rho F^{\,\lambda\nu} -
b^\nu b_\lambda \partial_\rho F^{\,\lambda\rho}\right) \nonumber\\
&& +\ m^2_\gamma\,A^\nu -\ \zeta\,b_\lambda\widetilde F^{\,\nu\lambda}\ =\
\partial^{\,\nu} B \,,
\label{2.2a}\\
&&\partial_\nu A^\nu =0\,.
\label{2.2b}
\end{eqnarray}
After contraction of eq.~(\ref{2.2a}) with $\partial_\nu$ we find
\begin{equation}
\partial^{\,2}B(x)=0\,,
\label{2.3}
\end{equation}
whence it follows that the auxiliary field is a decoupled
massless scalar field, which is not involved in dynamics.

After using eq.~(\ref{2.2b})
one can rewrite the field equations in terms of the gauge potential, {\it i.e.},
\ba
&&\left(1+ \xi\,\frac{b^{\,2}}{m_e^2}\right) \partial^{\,2} A^\nu
- (\,{\xi}/{m_e^2})\  \Bigl[\,(b\cdot \partial)^2 A^\nu - \partial^{\,\nu} (b\cdot \partial)(b\cdot A)
+ b^\nu \partial^{\,2} (b\cdot A)\,\Bigr] \nonumber\\
&& +\ \ m^2_{\,\gamma}\,A^\nu -\
\zeta\,\epsilon^{\,\nu\lambda\rho\sigma}\, b_\lambda \partial_\rho A_\sigma\
=\ 0\,.
\label{eq-4}
\ea
After contraction of eq.~(\ref{eq-4}) with $b^{\,\nu}$ we get
\begin{equation}
\Bigl(\partial^{\,2} + m^2_\gamma\Bigr) (b\cdot A)=0\,,
\label{2.4}
\end{equation}
for the special component $b\cdot A$ of the vector potential.
Thus, for this polarization, we actually find
the ordinary dispersion law of a real
massive scalar field,
whereas the two further components with polarizations orthogonal
to both $k_\nu$ and $b_\nu$ are affected by the fermion induced
LIV radiative corrections.

Going to the momentum representation,
the equations of motion take the form
\begin{eqnarray}
 K^{\,\nu\sigma}\,{\rm A}_\sigma(k)=0\,,\qquad k^\sigma{\rm A}_\sigma(k) = 0\,,\label{2.7}
\end{eqnarray}
where we have set
\begin{eqnarray}
K^{\,\nu\sigma}&\equiv&
(k^2 - m^2_\gamma)\,g^{\nu\sigma} - k^\nu k^\sigma -
\xi\,(\mbox{\tt D}/m_e^2)\,\mbox{\rm e}^{\nu\sigma}
+ i\,\zeta\,\epsilon^{\,\nu\lambda\rho\sigma}\, b_\lambda
k_\rho\,.
\end{eqnarray}
In order to pick out the two independent field degrees of freedom, we have
introduced the quantity
\begin{equation}
\mbox{\tt D}\equiv (b\cdot k)^2- b^2k^2
\label{2.8}
\end{equation}
and  the projector onto the two--dimensional hyperplane
orthogonal to $b_\nu$ and $k_\nu\,$ ,,
\begin{equation}
\mbox{\rm e}^{\nu\sigma}\equiv g^{\nu\sigma}-\frac{b\cdot k}{\mbox{\tt
D}}\left(b^\nu k^\sigma +b^\sigma k^\nu\right)+\frac{k^2}{\mbox{\tt
D}}b^\nu b^\sigma + \frac{b^2}{\mbox{\tt D}}k^\nu k^\sigma\,.
\label{2.9}
\end{equation}
One can always select two real orthonormal four--vectors corresponding to the
linear polarizations
in such a way that
\begin{equation}
\mbox{\rm e}_{\nu\sigma}=
-\sum_{a=1,2}\mbox{\rm e}^{(a)}_\nu\mbox{\rm e}^{(a)}_\sigma\,,\qquad
g^{\nu\sigma}\mbox{\rm e}^{(a)}_\nu\mbox{\rm e}^{(b)}_\sigma= -\,\delta^{ab}\,.
\label{2.11}
\end{equation}
It is also convenient to define another couple of four--vectors,
in order to describe the left-- and right--handed polarizations:
in our case, those generalize the circular polarizations of the
conventional QED. To this aim, let us first define
\begin{equation}
\epsilon^{\,\nu\sigma}\equiv \mbox{\tt
D}^{-1/2}\,\epsilon^{\,\nu\lambda\rho\sigma} b_\lambda
k_\rho\,.
\label{2.12}
\end{equation}
Notice that we can always choose $\mbox{\rm e}_\lambda^{(a)}$ to satisfy
\begin{equation}
\epsilon^{\,\nu\sigma}\mbox{\rm e}_\sigma^{(1)}=\mbox{\rm e}^{(2)\nu}\,,\qquad
\epsilon^{\,\nu\sigma}\mbox{\rm e}_\sigma^{(2)}=\ -\ \mbox{\rm e}^{(1)\nu} \,.
\label{2.14}
\end{equation}
Let us now construct the two orthogonal projectors
\begin{equation}
P_{\nu\sigma}^{(\pm)}\equiv{\frac12}\left(\mbox{\rm e}_{\nu\sigma}\pm
i\epsilon_{\nu\sigma}\right).
\label{2.15}
\end{equation}
and set, {\it e.g.},
\begin{eqnarray}
\varepsilon_\nu^{(L)}&\equiv& {\frac12}\left(\mbox{\rm e}_\nu^{(1)}+i\,\mbox{\rm
e}_\nu^{(2)}\right)=P_{\nu\sigma}^{(+)}\,\mbox{\rm e}^{(1)\,\sigma}\,,
\label{2.16a}\\
\varepsilon_\nu^{(R)}&\equiv& {\frac12}\left(\mbox{\rm
e}_\nu^{(1)}-i\,\mbox{\rm e}_\nu^{(2)}\right) =
P_{\nu\sigma}^{(-)}\,\mbox{\rm e}^{(1)\,\sigma}\,.
\label{2.16b}
\end{eqnarray}
We remind that, actually,  the left-- and right--handed (or
chiral) polarizations only approximately \cite{AGS} correspond to the circular
ones of Maxwell QED. In the presence of the CS kinetic term,
the field strengths of electromagnetic waves are typically not orthogonal
to the wave vectors.

Once the physical meaning of polarizations has been suitably focused,
one can readily
find the expression of the dispersion relations for the doubly transversal
photon modes ,
\begin{eqnarray}
&&\Biggl\{k^2 - \frac{\xi}{m_e^2}\Bigl[\,(b\cdot k)^2- b^2k^2\,\Bigr] -
m^2_\gamma \Biggr\}^2\, -\ \zeta^2 \Bigl[\,(b\cdot k)^2-b^2k^2\,\Bigr]=0\,.
\label{disp}
\end{eqnarray}
Evidently real solutions exist only iff
$$
\mbox{\tt D} = (b\cdot k)^2-b^2k^2 \geq 0\ ,
$$
consistently with our previous notations.
Notice that on the photon mass shell, deviations
off the light--cone are of order $|\,b_\nu|^{\,2}\,.$  As a consequence, the
on--shell momentum dependence of the polarization tensor (\ref{4.8}) is dominated by the
lowest order $k^2 = 0\,,$ whereas the higher orders in $k^2$ do represent
simultaneously higher orders in $b_\nu\,,$ which are neglected in the present
analysis.
\section{LIVQED consistency}
It is worthwhile to recall that
a time--like axial--vector $b^{\,\mu}$ is required
for a consistent fermion quantization \cite{AGS,adam,koslehn}.
Nonetheless,
one has to take into account that, on the one hand,
in the lack of a bare photon mass
and/or a bare CS vector of different direction a time--like vector $b^{\,\mu}$
just leads to a tachyonic massive photon \cite{ASS,adam}
and instability of the photodynamics, that means
imaginary energies for the soft photons.
On the other hand, a space--like vector $b^{\,\mu}$ causes problems for
fermion quantization \cite{AGS,koslehn} and {\it a fortiori}
for the very meaning of the radiative
corrections.  There are two ways to avoid this obstruction.

\noindent
A)\qquad If we adopt classical photons in Lorentz invariant QED to be massless $\mu_\gamma = 0$
then, in  such a situation, a fully induced LIV appears to be flawless only
for {\sl light--like} axial--vectors $b^{\,\mu}\,.$
In particular, for a light--like universal axial--vector
$b^{\,\mu}=(|\,\vec b\,|,\vec b\,)\,,$ we find the dispersion relations
for the LIV 1--particle states of a fermion species $f$ that read
\begin{eqnarray}
p_+^0+|\,\vec b\,|=\pm\sqrt{\Big(\vec p+\vec b\,\Big)^2+m_f^2}\,,\\
p_-^0-|\,\vec b\,|=\pm\sqrt{\Big(\vec p-\vec b\,\Big)^2+m_f^2}\,.
\end{eqnarray}
Now, it turns out that the requirement $p_\pm^2>0$ for the LIV free
1--particle spinor physical states just drives to
the high momenta cut--off $|\,\vec p\,|\le m_e^2/4\,|\,\vec b\,|$
which is well compatible with the
LIVDRS treatment of fermion loops.

Then one can use the induced values of the LIV parameters \gl{zeta},
\gl{param} in the case $b^2 = 0$ and the dispersion law for photons
(\ref{disp}) is reduced to
\begin{eqnarray}
k^2 - (\xi/m_e^2)\,(b\cdot k)^2 \pm \zeta\,b\cdot k=0\, .
\label{disp1}
\end{eqnarray}
For photon momenta $|\,\vec k\,|\gg |\,\vec b\,|$ one  approximately
finds the relationship for positive energies (frequencies)
\ba
&& k_0 \simeq |\,\vec k\,| (1 + \delta c_\theta) \mp
\zeta\,|\,\vec b\,|\,\sin^2\theta/2\,,\nonumber\\
&& \cos\theta \equiv \frac{{\vec b}\cdot \vec k}
{|\,\vec b\,|\,|\,\vec k\,|}\,,
\quad \delta c_\theta\equiv
\frac{2\xi}{m_e^2}\,|\,{\vec b}\,|^2\,\sin^4\theta/2 , \label{lowfreq}
\ea
and a similar expression for negative energies (frequencies).
One can see clearly that LIV entails an increase $\delta c_\theta$
of the light velocity, which makes it different from its
decrement generated by quantum gravity in the leading order \cite{quangrav,alfaro} .
Both the variation in the light velocity and the birefringence effect \cite{1}
caused by a phase shift between left-- and
right--polarized photons -- alternate signs in \gl{lowfreq} --
depend upon the direction of the wave vector $\vec k\,.$
Both effects do vanish in the  direction collinear with $\vec b\,.$
Thus the compilation of the UHECR data in search for deviations of the speed of light  must take
into account this possible anisotropy of photon spectra.

This is also true for the compilation
of the data on polarization plane rotation for radio waves from remote galaxies.
The earlier search for  this effect \cite{1,nod,0601095} led to the very stringent upper bound on values of
$|\,\vec b\,|\,<  \times 10^{-31}$ eV.
However, in addition to the previous remark on the
photon spectrum anisotropy, we would like to give more arguments
in favor of a less narrow
room for the possibility of LIV and CPT breaking in the Universe.
Indeed one must also take into account the apparent time variation of an anisotropic
CS vector , when its origin derives from the v.e.v.
of a parity--odd quintessence field  \cite{carr} very weakly coupled to photons.
That v.e.v. may well depend on time and obtain a tiny but sizeable
value in the later epoch of the Universe evolution \cite{ratra}, just like the cosmological constant
\cite{pad} might get. As well a non-vanishing CS vector may be induced also by the non-vanishing v.e.v.
of a dark matter component if its coupling to gravity is CPT odd . 
Eventually it means that, for large distances
corresponding to  earlier epochs in the Universe, one may not at all
experience this kind of LIV and CPT breaking.
Conversely, in a later time such a CS term may gradually rise up.
Then, the earlier radio sources -- galaxies and quasars with larger Hubble
parameters --  may not give any observable signal of birefringence, whereas the
individual evidences from a nearest radio source may be of a better confidence.
So far  we cannot firmly predict  on what is an actual
age of such CPT odd effects and therefore, to be conservative,
one has to rely upon the lab experiments and
meantime pay attention to the data from quasars of the nearest Universe.
Thus one may certainly trust to the estimations \cite{shore} performed in
the laboratory and the nearest Universe observations.
So far the most conservative value of the LIV parameter from
\cite{shore,DP} arises from hydrogen maser experiments:
namely, $|\,\vec b_e\,|<10^{-18}$ eV for electrons.

\medskip\noindent
B)\qquad Another way to implement the LIV, solely
by fermion coupling to an axial--vector background, is to start
with the Maxwell's photodynamics
supplemented by a bare and {\sl Lorentz invariant} photon mass $\mu_\gamma\,,$
so that
\be
m^2_\gamma = -\frac{2\alpha}{3\pi}\sum_f q_f^2\,b_f^2 + \mu^2_\gamma\,.
\ee
Then, for a genuine  time--like
$b^{\,\mu} = (\sum_f q_f\,b^0_f\,, 0,0,0)$ one finds
from eq.~\gl{disp} and the definition \gl{regodd} the following dispersion laws:
namely,
\ba
k_0^2 = \left(1 + \frac{\xi\,b_0^2}{m_e^2}\right)
\left(|\,\vec k\,|\,\pm \frac12\,\zeta\,b_0\right)^2
 + m^2_\gamma - b_0^2\Big\{\frac14\,\zeta^2 +
O(b_0\,|\,\vec k\,|/m^2_e)\Big\}\,.
\ea
Hence, if $ m_\gamma \geq \zeta\,b_0/ 2= 8 \alpha\,b_0/\pi\,,$ then the photon energy
keeps real for any wave vector $\vec k$  and LIVQED happens to be consistent. Meantime the longitudinal photon
polarization exhibits the entire mass $m_{\,\gamma} $.  
Then the present day very stringent experimental bound
on the photon mass \cite{PDG}, $m_\gamma < 6\times 10^{-17}$ eV,
does produce the  limit $b_0 <
3\times  10^{-15}\ {\rm eV}\, .$. 

\medskip
To conclude we would like to make few more comments on estimates for the LIV vector components.\\
1) There are no better bounds on $b_\mu$
coming from the UHECR
data on the speed of light for photons. This is because the increase of the
speed of light depends quadratically on components of $b_\mu\,.$ Thus,
for example, the data cited in \cite{2,shore} do imply less severe
bounds on $|\,\vec{b}\,|$ or $b_0$ than those ones above mentioned.\\
2)  For the LIVQED examined in the present paper, the typical bounds on LIV and CPT breaking
parameters in the context of quantum gravity phenomenology
are not good enough to compete with the laboratory estimations.
They are, in fact,   of a similar order of
magnitude as other LIV effects in the high energy astrophysics.\\
3) An interesting bound on deviations of the speed of light is given in \cite{lieu}
where, in the spirit of quantum gravity  phenomenology, space--time
fluctuations are addressed to produce modifications of the speed of light and,
as well, of the photon dispersion relations exhibiting helicity dependent effects.
Using an interferometric technique,  the authors of ref.~\cite{lieu}
were able to estimate $\Delta c < 10^{-32}\,.$  However this estimation
does not imply a better bound for a LIV vector $b_\mu\,,$ as it actually gives
$b_0 < 10^{-12}\ {\rm eV}\,,$ which is certainly in
agreement with the  more stringent bounds discussed above .

\section*{Acknowledgments}
A.A.A. was supported by RFBR, grant 05-02-17477 and  by the Programs
RNP 2.1.1.1112 and LSS-5538.2006.2.  P.G. and R.S. were supported by INFN, grant IS/PI13.
J.A. wants to thank the kind hospitality of Universidad Aut\'onoma de Madrid
where he spent a Sabbatical year using grant SAB2003-0238 of 
Secretaria de Estado de Universidades e Investigaci\'on(Spain).The work of J.A. has been partially
supported by Fondecyt 1060646.
 M.C. was partially supported by DIPUC and also wishes to thank ICTP for the
kind hospitality during his visit in the framework of the Federation Scheme.


\end{document}